\documentstyle[12pt,epsfig,amsfonts]{article}
\title{Self force on a point-like source coupled with\\ massive scalar field}
\author{Yurij Yaremko\footnote{Electronic mail: yar@ph.icmp.lviv.ua}}
\date{\it Institute for Condensed Matter Physics, \\
1 Svientsitskii St., 79011 Lviv, Ukraine}
\begin{document}

\maketitle
\begin{abstract}
The problem of determining the radiation reaction force experienced by a  
scalar charge moving in flat spacetime is investigated. A consistent 
renormalization procedure is used, which exploits the Poincar\'e invariance 
of the theory. Radiative parts of Noether quantities carried by massive 
scalar field are extracted. Energy-momentum and angular momentum balance 
equations yield Harish-Chandra equation of motion of radiating 
charge under the influence of an external force. This equation includes 
effect of particle's own field. The self force produces a time-changing 
inertial mass.
\end{abstract}



\section{Introduction}
\setcounter{equation}{0}
The problem of calculating the motion of an isolated point-like charge 
coupled to massive scalar field in flat spacetime is an old one which is 
currently receiving renewed interest. Classical equations of motion of a 
point particle interacting with a neutral massive vector field were first 
found by Bhabha \cite{Bha} following a method originally developed by Dirac 
\cite{Dir} for the case of electromagnetic field. In this method the finite 
force and self-force terms in the equations of motion are obtained from the 
conservation laws for the energy-momentum tensor of the field. It was 
extended by Bhabha and Harish-Chandra \cite{BHC} to particles 
interacting with any generalized wave field and was applied to the motion 
of a simple pole of massive scalar field by Harish-Chandra \cite{HC}. 

The principal new feature of the field which carries rest mass in addition 
to energy and momentum is that it is nonlocal. (The field depends not 
only on the current state of motion of the source but on its past history.) 
Physically it is due to the fact that the massive field propagates at 
all speeds smaller than the velocity of light. 

In \cite{H1,H2,Hv1,Hv2} a consistent theory of action at a distance was 
formulated in a case of point sources coupled to massive scalar or vector 
fields. In electrodynamics of Wheeler and Feynman \cite{WF}, the interaction 
is assumed to be symmetric in time. The symmetric case is the only one for 
which the equations of motion follow from a variational principle. However,
they do not contain any terms describing radiation damping. Such terms do 
appear if the assumption of complete absorption is applied to these 
equations. By this the authors \cite{WF} mean that the total advanced field 
of all the particles in the Universe equals their total retarded field. 

The equations of motion obtained from the field-theoretical and 
action-at-a-distance point of view are different: the integrals over entire 
world line of the particle substitute for integrals over the past motion 
which appear in case of purely retarded fields \cite{Bha,HC}. In 
\cite{HM,CHv,Hv3} the total cross sections for the scattering of the 
various kinds of mesons by a heavy particle (nucleon) were 
calculated\footnote{In the present paper we shall not identify neither 
fields nor sources with any currently known particles.} and compared with 
those obtained within the  Bhabha and Harish-Chandra approach. The 
predictions following from the two approaches should be exploited to 
furnish an experimental decision between the two theories. It is worth 
noting that in case of the retarded interactions, Havas \cite{H1} and
Crownfield and Havas \cite{HvC} obtain the Bhabha \cite{Bha} and 
Harish-Chandra \cite{HC} equations. 

In the present paper we calculate energy-momentum and angular momentum 
carried by outgoing massive scalar waves. Effective equations of motion of 
radiating scalar source will be obtained via the consideration of 
energy-momentum and angular momentum balance equations. The conservation 
laws are an immovable fulcrum about which tips the balance of truth 
regarding renormalization and radiation reaction. The verification is not a 
trivial matter, since the Klein-Gordon field generated by the scalar 
charge holds energy near the particle. This circumstance makes the procedure 
of decomposition of the Noether quantities into bound and radiative parts 
unclear.

In \cite{CM,Cw69,Cw70} Cawley and Marx study the massive scalar radiation 
from a point source with a prescribed world line. The authors assume that 
the particle accelerates only over a portion of the world line which 
corresponds to a finite proper time interval. They evaluate the 
energy-momentum which flows across a fixed three-dimensional sphere of large 
radius $R$. The radiation part of energy-momentum carried by massive scalar 
field was extracted which depends on $R$ explicitly. It casts serious doubt 
on the validity of the result. In the present paper we apply a consistent 
splitting procedure which obeys the spirit of Dirac scheme of 
decomposition of electromagnetic potential into singular (symmetric) and 
regular (radiative) components \cite{Dir}.

Recently \cite{Q}, Quinn has obtained an expression for the self-force on a 
point-like particle coupled to a massless scalar field arbitrarily moving in 
a curved spacetime. It is worth noting that in curved background massless 
waves propagate not just at speed of light, but at all speeds smaller than 
or equal to the speed of light. (It can be understood as the result of 
interaction between the radiation and the spacetime curvature.) 
Therefore, the particle may ``fill'' its own field, which will act on it 
just like an external field. In \cite{Q99} Quinn establishes that the total 
work done by the scalar self-force matches the amount of energy radiated 
away by the particle.

Using Quinn's general expression, Pfenning and Poisson \cite{PPs} calculate 
the self-force experienced by a point scalar charge moving in a weakly 
curved spacetime. It is characterized by a generic Newtonian potential 
$\Phi$ which determines the small deviation of the metric $g_{\alpha\beta}$ 
with respect to the Minkowski values $\eta_{\alpha\beta}={\rm 
diag}(-1,1,1,1)$. Potential $\Phi$ behaves as $-M/r$ at large distances 
$r$ from the bounded mass distribution of total mass $M$. In contrast to the 
electromagnetic case, the equations of motion of scalar charge does not 
provide conservation of the rest mass (see also \cite{Q,Q99}). In 
Refs.\cite{BHPs,HPs} this phenomenon is studied for various kinds of 
cosmological spacetimes.

In this paper we consider the radiation reaction problem for a point 
particle that acts as a source for a massive scalar field in Minkowski 
spacetime. It is organized as follows. In Section \ref{PF} we recall the 
Green's functions associated with the Klein-Gordon wave equation. 
Convolving them with the point-like source, we derive the retarded scalar 
potential and field strengths as well as their advanced counterparts. In 
Section \ref{Reg} we decompose the momentum 4-vector carried by massive 
scalar field into singular and regular parts. All diverging terms have 
disappeared into the procedure of mass renormalization while radiative 
terms survive. In analogous way we analyze the angular momentum of the 
Klein-Gordon scalar field. The radiative parts of Noether quantities 
carried by field and already renormalized particle's individual momentum 
and angular momentum constitute the total energy-momentum and total 
angular momentum of our particle plus field system. In Section \ref{meq} 
we derive the effective equations of motion of radiating scalar charge via 
analysis of balance equations. We show that it coincides with the 
Harish-Chandra equation \cite{HC}. In Section \ref{Concl} we discuss the 
result and its implications.

\section{Scalar potential and field strengths of a point-like 
scalar charge}\label{PF}
\setcounter{equation}{0}
The dynamics of a point-like charge coupled to massive scalar field is 
governed by the action \cite{Kos,Pois}
\begin{equation}\label{Itot}
I_{\rm total}=I_{\rm part}+I_{\rm int}+I_{\rm field}.
\end{equation}
Here
\begin{equation}
I_{\rm field}=-\frac{1}{8\pi}\int{\rm d}^4y\left(
\eta^{\alpha\beta}\varphi_\alpha\varphi_\beta+k_0^2\varphi^2\right)
\end{equation}
is an action functional for a massive scalar field $\varphi$ in flat 
spacetime. We shall use the metric tensor $\eta^{\alpha\beta}={\rm 
diag}(-1,1,1,1)$ and its inverse $\eta_{\alpha\beta}={\rm 
diag}(-1,1,1,1)$ to raise and lower indices, respectively. The mass 
parameter $k_0$ is a constant with the dimension of reciprocal length. The 
integration is performed over all the spacetime. The particle action is
\begin{equation}\label{SS}
I_{\rm part}=-m_0\int{\rm d}\tau\sqrt{-{\dot z}^2}
\end{equation}
where $m_0$ is the bare mass of the particle which moves on a world line 
$\zeta:{\mathbb R}\to{\mathbb M}_{\,4}$ described by relations 
$z^\alpha(\tau)$ which give the particle's coordinates as functions of 
proper time; ${\dot z}^\alpha(\tau)={\rm d} z^\alpha(\tau)/{\rm d}\tau$. 
Finally, the interaction term is given by
\begin{equation}
I_{\rm int}=g\int{\rm d}\tau\sqrt{-{\dot z}^2}\varphi(z)
\end{equation}
where $g$ is scalar charge carried by a four-dimensional Dirac distribution 
supported on $\zeta$: charge's density is zero everywhere, except at the 
particle's position where it is infinite. 

The action (\ref{Itot}) is invariant under infinitesimal transformations 
(translations and rotations) which constitute the Poincar\'e group. 
According to Noether's theorem, these symmetry properties yield 
conservation laws, i.e. those quantities that do not change with time. 

Variation on field variable $\varphi$ of action (\ref{Itot}) yields 
the Klein-Gordon wave equation
\begin{equation}\label{KG}
\left(\square -k_0^2\right)\varphi (y)=-4\pi\rho (y),
\end{equation}
where $\square=\eta^{\alpha\beta}\partial_\alpha\partial_\beta$ is the 
D'Alembert operator. We consider a scalar field satisfying eq.(\ref{KG}) in 
Minkowski spacetime with a point particle source
\begin{equation}\label{rd}
\rho(y)=g\int_{-\infty}^{+\infty}d\tau \delta^{(4)}(y-z(\tau)).
\end{equation}
A solution to eq.(\ref{KG}) can be expressed as
\begin{equation}\label{scf}
\varphi (y)=\int d^4xG(y,x)\rho (x).
\end{equation}
The relevant wave equation for the Green's function $G(y,x)$ is
\begin{equation}\label{Gwe}
\left(\square -k_0^2\right)G(y,x)=-4\pi\delta^{(4)}(y-x),
\end{equation}
where $\delta^{(4)}(y-x)$ is a four-dimensional Dirac functional in 
${\mathbb M}_4$. The retarded Greens function \cite{Bha,Kos,Pois}
\begin{equation}\label{Green}
G^{\rm ret}(y,x)=\theta(y^0-x^0)\left[
\delta(\sigma)-
\frac{k_0}{\sqrt{-2\sigma}}J_1(k_0\sqrt{-2\sigma})\theta(-\sigma)
\right]
\end{equation}
consists of singular part (this proportional to $\delta(\sigma)$) and smooth 
part (that proportional to $\theta(-\sigma)$). The former possesses support 
only on the past light cone of the field point $y$ while the latter 
represents a function supported within the past light cone of $y$. By 
$\sigma$ we denote Synge's world function in flat space-time \cite{Pois}
\begin{equation}\label{Sng}
\sigma(y,x)=\frac12\eta_{\alpha\beta}(y^\alpha -x^\alpha)(y^\beta 
-x^\beta)
\end{equation}
which is equal to half of the squared length of the geodesic connecting
two points in ${\mathbb M}_4$, namely ``base point'' $x$ and ``field point''
$y$. $\theta(y^0-x^0)$ is step function defined to be one if $y^0>x^0$, and 
defined to be zero otherwise, so that $G^{\rm ret}(y,x)$ vanishes in the 
past of $x$. $\theta(-\sigma)$ is the step function of $-\sigma(y,x)$ and 
$J_1$ is the first order Bessel's function of $k_0\sqrt{-2\sigma}$.

We substitute eq.(\ref{rd}) for the scalar density $\rho(x)$ in the 
right-hand side of eq.(\ref{scf}). Massive scalar waves propagate at all 
speeds smaller than or equal to the speed of light. Hence the retarded 
potential at each point $y$ of Minkowski space ${\mathbb M}_4$ consists of a 
local term as well as non-local one. The local term is evaluated at the 
retarded instant $\tau^{\rm ret}(y)$ which is determined by the intersection 
of the world line with the past cone of the field point $y$. The non-local 
term defines contribution from cone's interior. It reflect the circumstance 
that the retarded field at $y$ is generated also by the point source during 
its history prior $\tau^{\rm ret}(y)$.

Convolving the retarded Green's function (\ref{Green}) with the charge 
density (\ref{rd}) we construct the massive scalar field \cite{HC,H1,Pois}:
\begin{equation}\label{sg-sm}
\varphi^{\rm ret}(y)=\frac{g}{\sf 
r}-g\int\limits_{-\infty}^{\tau^{\rm ret}(y)} {\rm d}\tau
\frac{k_0J_1[k_0\sqrt{-(K\cdot K)}]}{\sqrt{-(K\cdot K)}}
\end{equation}
where $J_1$ is the first order Bessel's function of 
$k_0\sqrt{-2\sigma}$ which is rewritten as $k_0\sqrt{-(K\cdot K)}$. By 
$K^\mu=y^\mu - z^\mu(\tau)$ we denote the unique timelike (or null) vector 
pointing from the emission point $z(\tau)\in\zeta$ to a field point 
$y\in{\mathbb M}_4$. The upper limit
of the integral is the root of algebraic equation $\sigma(y,z(\tau))=0$ 
which satisfies causality condition $y^0-z^0(\tau^{\rm ret})>0$. By ${\sf 
r}$ we mean the retarded distance 
\begin{equation}\label{rtd}
{\sf 
r}(y)=-\eta_{\alpha\beta}(y^\alpha-z^\alpha(\tau^{\rm 
ret}))u^\beta(\tau^{\rm ret}).
\end{equation}
Because the speed of light is set to unity, it is also the spatial distance 
between $z(\tau^{\rm ret})$ and $y$ as measured in this momentarily comoving 
Lorentz frame where 4-velocity $u^\beta(\tau^{\rm ret})=(1,0,0,0)$.

Scalar field strengths are given by the gradient of the potential 
(\ref{sg-sm}). Let us differentiate the local term. Because $y$ and 
$z(\tau^{\rm ret})$ are linked by the light-cone mapping, a change of field 
point $y$ generally comes with a change $\tau^{\rm ret}$. Suppose that $y$ 
is displaced to the new field point $y+\delta y$. The new emission point 
$z(\tau^{\rm ret}+\delta \tau^{\rm ret})$ satisfies the algebraic equation 
$\sigma(y+\delta y,z(\tau^{\rm ret}+\delta \tau^{\rm ret}))=0$. Expanding 
this to the first order of infinitesimal displacements $\delta y$ and 
$\delta \tau^{\rm ret}$, we obtain $K_\alpha\delta y^\alpha+r\delta 
\tau^{\rm ret}=0$, or
\begin{equation}
\frac{\partial \tau^{\rm ret}}{\partial y^\alpha}=-\frac{K_\alpha}{{\sf 
r}(y)}.
\end{equation}
This relation allows us to differentiate the retarded distance (\ref{rtd}) 
in the local Coulomb-like term involved in eq.(\ref{sg-sm}).

Now we differentiate the non-local term in the potential (\ref{sg-sm}). 
Apart from the integral
\begin{equation}\label{fth}
f_\mu^{(\theta)}=g\int\limits_{-\infty}^{\tau^{\rm ret}(y)} {\rm d}\tau k_0^2
\frac{{\rm d}}{{\rm d}\Xi}\left(\frac{J_1(\Xi)}{\Xi}\right)k_0
\frac{K_\mu}{\sqrt{-(K\cdot K)}}
\end{equation}
the gradient $f_{{\rm tail},\mu}=f_\mu^{(\theta)}+f_\mu^{(\delta)}$ 
contains also local term
\begin{equation}\label{fdl}
f_\mu^{(\delta)}=gk_0^2\left.\frac{J_1(\Xi)}{\Xi}
\frac{K_\mu}{r}\right|_{\tau=\tau^{\rm ret}}
\end{equation}
which is due to time-dependent upper limit of integral in eq.(\ref{sg-sm}). 
Because of asymptotic behaviour of the first order Bessel's function with 
argument $\Xi:=k_0\sqrt{-(K\cdot K)}$ the local term $f_\mu^{(\delta)}$ is 
finite on the light cone where $\Xi=0$. It diverges on the particle's 
trajectory only.

To simplify the non-local contribution us much us possible we use the 
identity
\begin{equation}
\frac{k_0}{\sqrt{-(K\cdot K)}}=\frac{1}{(K\cdot u)}
\frac{{\rm d}\Xi}{{\rm d}\tau}
\end{equation}
in the integral (\ref{fth}) and perform integration by parts. On 
rearrangement, we add it to the expression (\ref{fdl}). The term which 
depends on the end points only annuls $f_\mu^{(\delta)}$. Finally, the 
gradient of potential (\ref{sg-sm}) becomes
\begin{equation}\label{grad}
\frac{\partial \varphi^{\rm ret}(y)}{\partial y^\mu}=-g\frac{1+(K\cdot 
a)}{{\sf r}^3}K_\mu+g\frac{u_\mu}{{\sf r}^2} + 
g\int\limits_{-\infty}^{\tau^{\rm ret}(y)} {\rm d}\tau 
k_0^2\frac{J_1(\Xi)}{\Xi}\left[\frac{1+(K\cdot a)}{r^2}K_\mu-
\frac{u_\mu}{r}\right]
\end{equation}
where $\Xi:=k_0\sqrt{-(K\cdot K)}$. As it is in the potential itself, 
particle's position, velocity, and acceleration in the local part are 
referred to the retarded instant $\tau^{\rm ret}(y)$ while ones under the 
integral sign are evaluated at instant $\tau\le \tau^{\rm ret}(y)$. The 
non-local part arises from source contributions interior to the light cone. 
This part of field is called the ``tail term''. The invariant quantity
\begin{equation}\label{rint}
r=-(K\cdot u)
\end{equation}
is an affine parameter on the time-like (null) geodesic that links $y$ to 
$z(\tau)$; it can be loosely interpreted as the time delay between $y$ and
$z(\tau)$ as measured by an observer moving with the particle.

The advanced Green's function is non-zero in the past of emission point $x$:
\begin{equation}\label{adv_Gr}
G^{\rm adv}(y,x)=\theta(-y^0+x^0)\left[
\delta(\sigma)-
\frac{k_0}{\sqrt{-2\sigma}}J_1(k_0\sqrt{-2\sigma})\theta(-\sigma)
\right].
\end{equation}
The advanced force
\begin{equation}\label{adv_f}
\frac{\partial \varphi^{\rm adv}(y)}{\partial y^\mu}=-g\frac{1+(K\cdot 
a)}{{\sf r}^3}K_\mu+g\frac{u_\mu}{{\sf r}^2} + 
g\int\limits^{+\infty}_{\tau^{\rm adv}(y)} {\rm d}\tau 
k_0^2\frac{J_1(\Xi)}{\Xi}\left[\frac{1+(K\cdot a)}{r^2}K_\mu-
\frac{u_\mu}{r}\right]
\end{equation}
is generated by the point charge during its entire future history following 
the advanced time associated with $y$. Particle's characteristics in the 
local part are referred to the instant $\tau^{\rm adv}(y)$.

\section{Bound and radiative parts of Noether quantities}\label{Reg}
\setcounter{equation}{0}
In this Section we decompose the energy-momentum and angular momentum 
carried by massive scalar field into the bound and radiative parts. The 
bound terms will be absorbed by particle's individual characteristics 
while the radiative terms exert the radiation reaction. We do not calculate 
the flows of the massive scalar field across a thin tube around a world line 
of the source. To extract the appropriate finite parts of energy-momentum 
and angular momentum we apply the scheme developed in Refs.\cite{Yar3D,Y3D}. 
In these papers the radiation reaction problem for an electric charge moving 
in flat spacetime of three dimensions is considered. A specific feature of 
$2+1$ electrodynamics is that both the electromagnetic potential and 
electromagnetic field are non-local: they depend not only on the current 
state of motion of the particle, but also on its past (or future) 
history. The scalar potential (\ref{sg-sm}) as well as the scalar field 
strengths (\ref{grad}) and (\ref{adv_f}) behave analogously.

Decomposition of Noether quantities into bound and radiative components 
satisfies the following conditions \cite{Yar3D,Y3D}:
\begin{itemize}
\item
proper non-accelerating limit of singular and regular parts;
\item
proper short-distance behaviour of regular part;
\item
Poincar\'e invariance and reparametrization invariance.
\end{itemize}
The first point means that in specific case of rectilinear uniform motion 
regular parts should vanish because of non-accelerating charge does not 
radiate. By ``proper short-distance behaviour'' we mean the finiteness of
integrand near the coincidence limit where point of emission placed on the 
world line tends to the field point which also lies on $\zeta$. (The bound 
parts of non-local conserved quantities in $2+1$ electrodynamics contain one 
integration over the world line while radiative ones are integrated over 
$\zeta$ twice.)

The scalar potential (\ref{sg-sm}) and the scalar field strengths 
(\ref{grad}) and (\ref{adv_f}) contain local terms as well as non-local 
ones. Local part of energy-momentum carried by massive scalar field is 
obtained in \cite{HC,H1,BV}. It is equal to one-half of the well-known 
Larmor rate of radiation integrated over the world line:
\begin{equation}\label{ploc}
p_{\rm loc,R}^\mu=\frac{g^2}{3}\int_{-\infty}^\tau{\rm d} sa^2(s)u^\mu(s).
\end{equation}
Similarly, the local part of radiated angular momentum is equal to the 
one-half of corresponding quantity in classical electrodynamics \cite{LV}:
\begin{equation}\label{pnloc}
M_{\rm loc,R}^{\mu\nu}=\frac{g^2}{3}\int_{-\infty}^\tau{\rm d} s
a^2_s\left[z^\mu_su^\nu_s-z^\nu_su^\mu_s\right]
+\frac{g^2}{3}\int_{-\infty}^\tau{\rm d} s
\left[u^\mu_sa^\nu_s-u^\nu_sa^\mu_s\right].
\end{equation}
There are singular terms associated with the Coulomb-like potential taken on 
particle's world line (see \ref{unmvd}, eq.(\ref{infin})). Inevitable 
infinity is absorbed by ``bare'' mass within the renormalization procedure. 

To find the ``tail'' parts of radiated Noether quantities sourced by the 
interior of the light cone we deal with the field defined on the world line 
only. Following the scheme presented in \cite{Yar3D,Y3D} we build our 
construction upon the tail part of the field strengths (\ref{grad}) 
evaluated at point $z(\tau_1)\in\zeta$: 
\begin{eqnarray}\label{tlF}
f_{{\rm 
tail},\mu}^{\rm 
ret}&=&\left.\frac{\partial\varphi^{\rm ret}_{\rm 
tail}(y)}{\partial y^\mu}\right|_{y=z(\tau_1)}\\
&=&g\int_{-\infty}^{\tau_1}{\rm d}\tau_2k_0^2\frac{J_1(\xi)}{\xi}
\left[\frac{1+(q\cdot a_2)}{r_2^2}q_\mu- \frac{u_{2,\mu}}{r_2}\right].
\nonumber
\end{eqnarray}
Here $q^\mu=z_1^\mu-z_2^\mu$ defines the unique timelike 4-vector pointing 
from an emission point $z(\tau_2)\in\zeta$ to a field point 
$z(\tau_1)\in\zeta$. Index $1$ indicates that particle's position, 
velocity, or acceleration is referred to the instant $\tau_1\in 
]-\infty,\tau]$ while index $2$ says that the particle's characteristics are 
evaluated at instant $\tau_2\le\tau_1$. We use the notations 
$\xi=k_0\sqrt{-(q\cdot q)}$ and $r_2=-(q\cdot u_2)$.

Next we consider the ``advanced'' counterpart of the expression (\ref{tlF}):
\begin{equation}\label{adF}
f_{{\rm 
tail},\mu}^{\rm 
adv}=g\int^{\tau}_{\tau_1}{\rm d}\tau_2k_0^2\frac{J_1(\xi)}{\xi}
\left[\frac{1+(q\cdot a_2)}{r_2^2}q_\mu- \frac{u_{2,\mu}}{r_2}\right].
\end{equation}
It is intimately connected with the gradient of $\varphi^{\rm adv}_{\rm 
tail}(y)$ evaluated at point $y=z(\tau_1)$. Note that the advanced force 
(\ref{adv_f}) is generated by the point charge during its entire 
future history. In (\ref{adF}) the domain of integration is the 
portion of the world line which corresponds to the interval 
$\tau_2\in[\tau_1,\tau]$ where $\tau$ is the so-called ``instant of 
observation''. This instant arise naturally in \cite{Yar3D,Y3D} where an 
interference of outgoing waves at the plane of constant value of $y^0$ is 
investigated. Its role is elucidated in \cite[Figs.2-4]{Yar3D} and 
\cite[Figs.1,2]{Y3D}.

We postulate that non-local part of energy-momentum carried by outgoing 
radiation is one-half of work done by the retarded tail force minus 
one-half of work performed by the advanced one, taken with opposite sign:
\begin{equation}
p^{\rm R}_{{\rm tail},\mu}=-\frac{g}{2}\left(
\int_{-\infty}^{\tau}{\rm d}\tau_1f_{{\rm tail},\mu}^{\rm ret}-
\int_{-\infty}^{\tau}{\rm d}\tau_1f_{{\rm tail},\mu}^{\rm adv}
\right).
\end{equation}
It is obvious that the ``advanced'' domain of integration, 
$\int_{-\infty}^\tau{\rm d}\tau_1\int_{\tau_1}^{\tau}{\rm d}\tau_2$, 
is equivalent to 
$\int_{-\infty}^\tau{\rm d}\tau_2\int^{\tau_2}_{-\infty}{\rm d}\tau_1$.
It can be replaced by the ``retarded'' one, $\int_{-\infty}^\tau{\rm d} 
\tau_1\int_{-\infty}^{\tau_1}{\rm d}\tau_2$, via interchanging of indices 
``first'' and ``second'' in the integrand. The ``tail'' part of 
energy-momentum carried by outgoing radiation becomes
\begin{equation}\label{p_nlrad} 
 p_{\rm tail,R}^\mu=\frac{g^2}{2}\int_{-\infty}^\tau{\rm d}\tau_1
\int_{-\infty}^{\tau_1}{\rm d}\tau_2
k_0^2\frac{J_1(\xi)}{\xi}\left[
-\frac{1+(q\cdot a_2)}{r_2^2}q^\mu+\frac{u_2^\mu}{r_2}
-\frac{1-(q\cdot a_1)}{r_1^2}q^\mu+\frac{u_1^\mu}{r_1}
\right]
\end{equation} 
where $r_a=-(q\cdot u_a)$. It is noteworthy that all the moments are before 
the observation instant $\tau$, and the retarded causality is not violated.

In the specific case of a uniformly moving source $q^\mu=u^\mu 
(\tau_1-\tau_2)$ and $r_a=\tau_1-\tau_2$ for both $a=1$ and $a=2$. Hence the 
bracketed integrands in eq. (\ref{p_nlrad}) is identically equal to zero. 
The local parts of radiation (\ref{ploc}) and (\ref{pnloc}) vanish if 
$u^\mu={\rm const}$. As could be expected, nonaccelerating scalar charge 
does not radiate.

Now we evaluate the short-distance behaviour of the expression under the 
double integrals in eq.(\ref{p_nlrad}). Let $\tau_1$ be fixed and 
$\tau_1-\tau_2:=\Delta$ be a small parameter. With a degree of accuracy 
sufficient for our purposes
\begin{eqnarray}\label{delta}
\sqrt{-(q\cdot q)}&=&\Delta\\
q^\mu&=&\Delta\left[u_1^\mu-a_1^\mu\frac{\Delta}{2}+
{\dot a}_1^\mu\frac{\Delta^2}{6}\right]\nonumber\\
u_2^\mu&=&u_1^\mu-a_1^\mu\Delta+{\dot a}_1^\mu\frac{\Delta^2}{2}.\nonumber
\end{eqnarray}
Substituting these into integrands of the double integrals of 
eq.(\ref{p_nlrad}) and passing to the limit $\Delta\to 0$ yields vanishing 
expression. Hence the subscript ``R'' stands for ``regular'' as well as for 
``radiative''.

In analogous way we construct the non-local part of radiated angular 
momentum. First of all we introduce the torque of the retarded tail force 
(\ref{tlF}) and its advanced counterpart:
\begin{equation}
 m_{{\rm tail},\mu\nu}^{\rm ret}=z_{1,\mu}f_{{\rm tail},\nu}^{\rm ret}-
z_{1,\nu}f_{{\rm tail},\mu}^{\rm ret},\qquad
m_{{\rm tail},\mu\nu}^{\rm adv}=z_{1,\mu}f_{{\rm tail},\nu}^{\rm adv}-
z_{1,\nu}f_{{\rm tail},\mu}^{\rm adv}.
\end{equation}
The desired expression is equal to the one-half of integral of 
$m_{{\rm tail},\mu\nu}^{\rm ret}$ over the world line up to observation 
instant $\tau$ minus one-half of integral of 
$m_{{\rm tail},\mu\nu}^{\rm adv}$, taken with opposite sign:
\begin{eqnarray}\label{M_nlrad} 
 M_{\rm tail,R}^{\mu\nu}=\frac{g^2}{2}
\int_{-\infty}^\tau{\rm d}\tau_1\int_{-\infty}^{\tau_1}{\rm d}\tau_2
k_0^2\frac{J_1(\xi)}{\xi}\left[
\frac{1+(q\cdot 
a_2)}{r_2^2}\left(z_1^\mu z_2^\nu -z_1^\nu z_2^\mu\right)
+\frac{z_1^\mu u_2^\nu-z_1^\nu u_2^\mu}{r_2}
\right.\nonumber\\
\left.
+\frac{1-(q\cdot 
a_1)}{r_1^2}\left(z_1^\mu z_2^\nu -z_1^\nu z_2^\mu\right)
+\frac{z_2^\mu u_1^\nu-z_2^\nu u_1^\mu}{r_1}
\right].
\end{eqnarray}
In the specific case of constant velocity this expression vanishes. 
Substituting eqs.(\ref{delta}) in the integrand passing to the limit 
$\Delta\to 0$ leads to zero. 

We postulate that bound part of energy-momentum carried by non-local part of 
massive scalar field is one-half of sum of work done by the retarded and the 
advanced tail forces:
\begin{eqnarray}\label{pS}
p^{\rm S}_{{\rm tail},\mu}(\tau)&=&
-\frac{g}{2}\left(
\int_{-\infty}^{\tau}{\rm d}\tau_1f_{{\rm tail},\mu}^{\rm ret}+
\int_{-\infty}^{\tau}{\rm d}\tau_1f_{{\rm tail},\mu}^{\rm adv}
\right)\\
&=&-\frac{g^2}{2}
\int_{-\infty}^\tau 
{\rm d} sk_0^2\frac{J_1(\xi)}{\xi}\frac{q_\mu(\tau,s)}{r_\tau}.\nonumber
\end{eqnarray}
The bound part of angular momentum also contains only one integration over 
the fragment of particle's world line:
\begin{eqnarray}\label{MS}
M^{\rm S}_{\rm tail,\mu\nu}(\tau)&=&
-\frac{g}{2}\left(
\int_{-\infty}^{\tau}{\rm d}\tau_1m_{{\rm tail},\mu\nu}^{\rm ret}+
\int_{-\infty}^{\tau}{\rm d}\tau_1m_{{\rm tail},\mu\nu}^{\rm adv}
\right)\\
&=&\frac{g^2}{2}
\int_{-\infty}^\tau {\rm d} sk_0^2\frac{J_1(\xi)}{\xi}
\frac{z_{\tau,\mu}z_{s,\nu}-z_{\tau,\nu}z_{s,\mu}}{r_\tau}.
\nonumber
\end{eqnarray}
Here index $\tau$ indicates that particle's position, velocity, or 
acceleration is referred to the observation instant $\tau$ while index 
$s$ says that the particle's characteristics are evaluated at instant 
$s\le\tau$. We denote $r_\tau=-(q\cdot u_\tau)$.

If $u^\mu={\rm const}$ that $\xi=k_0(\tau-s)$ and $q^\mu/r_\tau=u^\mu$. Since
\begin{equation}\label{b_int}
\int_{-\infty}^\tau{\rm d} s\frac{J_1[k_0(\tau -s)]}{\tau -s}=1,
\end{equation}
the field generated by a uniformly moving charge contributes an amount 
$p^\mu_{\rm tail,S}=-1/2g^2k_0u^\mu$ to its energy-momentum. This finding is 
in line with that of Appendix \ref{unmvd} where is established that if the 
particle is permanently at rest, the scalar meson field adds  $-1/2g^2k_0$ 
to its energy. The bound angular momentum is $M^{\mu\nu}_{\rm tail,S}=
z_0^\mu p^\nu_{\rm tail,S}-z_0^\nu p^\mu_{\rm tail,S}$ in case of uniform 
motion. We suppose, that the bound parts (\ref{pS}) and (\ref{MS}) of 
energy-momentum and angular momentum, respectively, are permanently 
``attached'' to the charge and are carried along with it. It is worth noting 
that they possess the proper short-distance behaviour and, therefore, do not 
diverge. The ``local'' Coulomb-like infinity is the only divergency stemming 
from the pointness of the source.

There is one more question to be answered: is the choice of the force 
(\ref{tlF}) the only one? If not there exists an alternative expression for 
radiated energy-momentum. It is interesting to apply our decomposition 
procedure to the massive scalar field as it is described in 
Refs.\cite{CM,Cw69,Cw70}. Cawley and Marx \cite{CM} remove the local 
Coulomb-like term from the retarded scalar potential. Using the recurrent 
relation $J_1(\Xi)=-{\rm d} J_0/{\rm d}\Xi$ between Bessel's function of 
order zero and of order one in eq.(\ref{sg-sm}) yields 
\begin{equation}\label{CMfi}
\varphi^{\rm ret}(y)=g\int\limits_{-\infty}^{\tau^{\rm ret}(y)} {\rm d}\tau
J_0(\Xi)\frac{1+(K\cdot a)}{r^2}
\end{equation}
after integration by parts. The authors state that the Klein-Gordon source 
does not emanate massless radiation. Following their approach, we rewrite 
the scalar field strengths (\ref{grad}) as follows:
\begin{equation}\label{CMgrad}
\frac{\partial \varphi^{\rm ret}(y)}{\partial y^\mu}= 
g\int\limits_{-\infty}^{\tau^{\rm ret}(y)} {\rm d}\tau J_0(\Xi)\left\{
-3\frac{\left[1+(K\cdot a)\right]^2}{r^4}K_\mu-
\frac{(K\cdot {\dot a})}{r^3}K_\mu
+3\frac{1+(K\cdot a)}{r^3}u_\mu
+\frac{a_\mu}{r^2}\right\}.
\end{equation}
Putting the field point $z(\tau_1)\in\zeta$ and the emission point 
$z(\tau_2)\in\zeta$, we obtain the scalar self-field: 
\begin{equation}\label{CMself}
F_\mu^{\rm ret}= 
g\int\limits_{-\infty}^{\tau_1} {\rm d}\tau_2 J_0(\xi)\left\{
-3\frac{\left[1+(q\cdot a_2)\right]^2}{r_2^4}q_\mu -
\frac{(q\cdot {\dot a}_2)}{r_2^3}q_\mu
+3\frac{1+(q\cdot a_2)}{r_2^3}u_{2,\mu} 
+\frac{a_{2,\mu}}{r_2^2}\right\}.
\end{equation}
Similarly one can construct its ``advanced'' counterpart which is generated 
by the point source during its history after $\tau_1$ up to the observation 
instant $\tau$.

Our next task is to extract the radiation part of energy-momentum carried by 
Cawley's scalar field (\ref{CMfi}). Since the radiation does not propagate 
with the speed of light, the Larmor-like term (\ref{ploc}) does not appear. 
The tail contribution to the radiation 
\begin{eqnarray}\label{pnl_rad} 
p^{\rm R}_\mu&=&-\frac{g}{2}\left(
\int_{-\infty}^{\tau}{\rm d}\tau_1F^{\rm ret}_\mu- 
\int_{-\infty}^{\tau}{\rm d}\tau_1F_\mu^{\rm adv}
\right)
\\
&=&-\frac{g^2}{2}
\int_{-\infty}^\tau{\rm d}\tau_1\int_{-\infty}^{\tau_1}{\rm d}\tau_2
k_0^2J_0(\xi)\left\{
-3\frac{\left[1+(q\cdot a_2)\right]^2}{r_2^4}q_\mu -
\frac{(q\cdot {\dot a}_2)}{r_2^3}q_\mu\right.\nonumber\\
&&\left. 
+3\frac{1+(q\cdot a_2)}{r_2^3}u_{2,\mu} 
+\frac{a_{2,\mu}}{r_2^2}-
\left[
3\frac{\left[1-(q\cdot a_1)\right]^2}{r_1^4}q_\mu +
\frac{(q\cdot {\dot a}_1)}{r_1^3}q_\mu\right.\right.\nonumber\\
&&\left.\left. 
-3\frac{1-(q\cdot a_1)}{r_1^3}u_{1,\mu} 
+\frac{a_{1,\mu}}{r_1^2}
\right]
\right\}\nonumber
\end{eqnarray} 
is meaningful only. Let us study the short-distance behaviour. Having 
inserted the relations (\ref{delta}), we see that the double integral is 
ill defined because the integrand diverges at the edge $\tau_2=\tau_1$ of 
the integration domain $D_\tau=\{(\tau_1,\tau_2)\in{\mathbb R}^{\,2}:
\tau_1\in]-\infty,\tau],\tau_2\le\tau_1\}$. It is because the Coulomb-like 
divergency moves under the integral sign (cf. eqs.(\ref{sg-sm}) and 
(\ref{CMfi})).

In the following Section we check the formula (\ref{p_nlrad}) and 
(\ref{M_nlrad}) via analysis of energy-momentum and angular momentum balance 
equations. Analogous equations yield correct equation of motion of radiating 
charge in conventional 3+1 electrodynamics \cite{Y02,Y03} as well as in six 
dimensions \cite{Y6D}. It is reasonable to expect that conservation laws 
result correct equation of motion of point-like source coupled with 
massive scalar field where radiation back reaction is taken into account. 

\section{Equation of motion of radiating charge}\label{meq}
\setcounter{equation}{0}

The equation of motion of radiating pole of massive scalar field was derived 
by Harish-Chandra \cite{HC} in 1946. (An alternative derivation was produced 
by Havas and Crownfield in \cite{HvC}.) Following the method of Dirac 
\cite{Dir}, Harish-Chandra enclosed the world line of the particle by a 
narrow tube, the radius of which will in the end be made to tend to zero. 
The author calculates the flow of energy and momentum out of the portion of 
the tube in presence of an external field. The condition was imposed that 
the flow depends only on the states at the two ends of the tube (the 
so-called ``inflow theorem'', see \cite{BHC44,BHC}). After integration over 
the tube along the world line and a limiting procedure, the equation of 
motion was derived. In our notation it looks as follows:
\begin{eqnarray}\label{HCme}
&&\!\!\!\!\!\!\!\!\!\!m_0a_\tau^\mu-\frac{g^2}{3}\left({\dot a}^\mu_\tau
-a^2_\tau u^\mu_\tau\right)-\frac{g^2}{2}k_0^2u_\tau^\mu
+g^2\int_{-\infty}^\tau {\rm d} sk_0^4\frac{J_2(\xi)}{\xi^2}q^\mu
+g^2\frac{{\rm d}}{{\rm d}\tau}\left(u_\tau^\mu\int_{-\infty}^\tau 
{\rm d} sk_0^2\frac{J_1(\xi)}{\xi}\right)\nonumber\\
&=&g\eta^{\mu\alpha}\frac{\partial \varphi_{\rm ext}}{\partial 
z^\alpha}+g\frac{{\rm d}}{{\rm d}\tau}\left(u_\tau^\mu \varphi_{\rm 
ext}\right)
\end{eqnarray}
where $m_0$ is an arbitrary constant identified with the mass of the 
particle and $\varphi_{\rm ext}$ is the scalar potential of the external 
field evaluated at the current position of the particle. $J_2(\xi)$ is the 
second order Bessel's function. In this Section the Harish-Chandra equation 
will be obtained via analysis of energy-momentum and angular momentum 
balance equations.

In previous Section we introduce the radiative part $p_{\rm R}=p_{\rm 
loc,R}+p_{\rm tail,R}$ of energy-momentum carried by the field. We proclaim 
that it alone exerts a force on the particle. We assume that the bound part, 
$p_{\rm S}$, is absorbed by particle's 4-momentum so that ``dressed'' 
charged particle would not undergo any additional radiation reaction. 
Already renormalized particle's individual four-momentum, say $p_{\rm 
part}$, together with $p_{\rm R}$ constitute the total energy-momentum of 
our composite particle plus field system:  $P=p_{\rm part}+p_{\rm R}$. We 
suppose that the gradient of the external potential matches the change of 
$P$ with time: 
\begin{eqnarray}\label{pdot}
{\dot p}^\mu_{\rm part}(\tau)&=&-{\dot p}^\mu_{\rm R}+
g\eta^{\mu\alpha}\frac{\partial \varphi_{\rm ext}}{\partial z^\alpha}\\
&=&-\frac{g^2}{3}a^2(\tau)u^\mu_\tau 
+\frac{g^2}{2}\int_{-\infty}^\tau{\rm d} s
k_0^2\frac{J_1(\xi)}{\xi}\left[
\frac{1+(q\cdot a_s)}{r_s^2}q^\mu-\frac{u_s^\mu}{r_s}
+\frac{1-(q\cdot a_\tau)}{r_\tau^2}q^\mu-\frac{u_\tau^\mu}{r_\tau}
\right]\nonumber\\
&+&g\eta^{\mu\alpha}\frac{\partial \varphi_{\rm ext}}{\partial 
z^\alpha}.\nonumber
\end{eqnarray} 
The overdot means the derivation with respect to proper time $\tau$.

Our next task is to derive expression which explain how three-momentum of 
``dressed'' charged particle depends on its individual characteristics 
(velocity, position, mass etc.). We do not make any assumptions about the 
particle structure, its charge distribution and its size. We only assume 
that the particle 4-momentum $p_{\rm part}$ is finite. To find out 
the desired expression we analyze conserved quantities corresponding to the 
invariance of the theory under proper homogeneous Lorentz transformations. 
The total angular momentum, say $M$, consists of particle's angular 
momentum $z\wedge p_{\rm part}$ and radiative part of angular momentum 
carried by massive scalar field:
\begin{equation}\label{Mtot}
M^{\mu\nu}=z_\tau^\mu p_{\rm part}^\nu(\tau) 
- z_\tau^\nu p_{\rm part}^\mu(\tau) + M^{\mu\nu}_{\rm R}(\tau).
\end{equation}

We assume that the torque $z_\tau^\mu \partial^\nu \varphi_{\rm 
ext}-z_\tau^\nu \partial^\mu \varphi_{\rm ext}$ of the potential external 
force matches the change of $M$ with time. Having differentiated 
(\ref{Mtot}) where the radiated angular momentum $M^{\mu\nu}_{\rm 
R}=M^{\mu\nu}_{\rm R,loc}+ M^{\mu\nu}_{\rm R,tail}$ is determined by 
eqs.(\ref{pnloc}) and (\ref{M_nlrad}), and inserting 
eq.(\ref{pdot}) we arrive at the equality
\begin{equation}\label{Mdot}
u_\tau\wedge \left(p_{\rm part}+\frac{g^2}{3}a_\tau+
\frac{g^2}{2}\int_{-\infty}^\tau {\rm d} sk_0^2\frac{J_1(\xi)}{\xi}
\frac{q}{r_\tau}\right)=0.
\end{equation}
Apart from usual velocity term, the 4-momentum of ``dressed'' particle
contains also a contribution from field:
\begin{equation}\label{ppart}
p_{\rm part}^\mu=mu_\tau^\mu-\frac{g^2}{3}a_\tau^\mu-
\frac{g^2}{2}\int_{-\infty}^\tau {\rm d} sk_0^2\frac{J_1(\xi)}{\xi}
\frac{q^\mu}{r_\tau}.
\end{equation}
The local part is the scalar analog of Teitelboim's expression \cite{Teit} 
for individual 4-momentum of a ``dressed'' electric charge in 
conventional electrodynamics. The integral term is then nothing but the 
bound part (\ref{pS}) of energy-momentum carried by the massive scalar 
field. 

The expression for the scalar function $m(\tau)$ is find in \ref{mass} via 
analysis of differential consequences of conservation laws. We derive that 
{\it already renormalized} dynamical mass $m$ depends on particle's 
evolution before the observation instant $\tau$:
\begin{equation}\label{mm}
m=m_0+g^2\int_{-\infty}^\tau {\rm d} 
sk_0^2\frac{J_1[\xi(\tau,s)]}{\xi(\tau,s)}-
g\varphi_{\rm ext}.
\end{equation} 
The constant $m_0$ can be identified with the renormalization constant 
in action (\ref{SS}) which absorbs Coulomb-like divergence stemming from 
local part of potential (\ref{sg-sm}). It is of great importance that the 
dynamical mass, $m$, will vary with time: the particle will necessarily gain 
or lost its mass as a result of interactions with its own field as well as 
with the external one. The field of a uniformly moving charge contributes an 
amount $g^2k_0$ to its inertial mass.

To derive the effective equation of motion of radiating charge we replace 
${\dot p}_{\rm part}^\mu$ in left-hand side of eq.(\ref{pdot}) by 
differential consequence of eq.(\ref{ppart}). We apply the formula
\begin{equation}\label{df}
\frac{\partial}{\partial\tau}\int_{-\infty}^\tau{\rm d} sf(\tau,s)=
\int_{-\infty}^\tau{\rm d} s\left(\frac{\partial f}{\partial\tau}+
\frac{\partial f}{\partial s}\right).
\end{equation} 
At the end of a straightforward calculations, we obtain 
\begin{equation}\label{me}
 ma^\mu_\tau+{\dot m}u^\mu_\tau=\frac{g^2}{3}\left({\dot a}^\mu_\tau
-a^2_\tau u^\mu_\tau\right)
+ g^2\int_{-\infty}^\tau {\rm d} sk_0^2\frac{J_1(\xi)}{\xi}\left[
\frac{1+(q\cdot a_s)}{r_s^2}q^\mu-\frac{u_s^\mu}{r_s}
\right]+g\eta^{\mu\alpha}\frac{\partial \varphi_{\rm ext}}{\partial 
z^\alpha}
\end{equation}
where dynamical mass $m(\tau)$ is defined by eq.(\ref{mm}). The local 
part of the self-force is one-half of well-known Abraham radiation reaction 
vector while the non-local one is then nothing but the tail part of 
particle's scalar field strengths (\ref{grad}) acting upon itself (see 
eq.(\ref{tlF})). Indeed, since the massive field does not propagate with the 
velocity of light, the charge may ``fill'' its own field, which will act on 
it just like an external field.

Now we compare this effective equation of motion with the Harish-Chandra 
equation (\ref{HCme}). The latter can be simplified substantially.
Having used the recurrent relation
\begin{equation}\label{rcJ}
J_2(\xi)=\frac{J_1(\xi)}{\xi}-\frac{{\rm d} J_1(\xi)}{{\rm d}\xi}
\end{equation}
between Bessel functions of order two and of order one, after integration by 
parts we obtain
\begin{equation}\label{rc}
 g^2\int_{-\infty}^\tau {\rm d} sk_0^4\frac{J_2(\xi)}{\xi^2}q^\mu=
\frac{g^2}{2}k_0^2u_\tau^\mu -
g^2\int_{-\infty}^\tau {\rm d} sk_0^2\frac{J_1(\xi)}{\xi}\left[
\frac{1+(q\cdot a_s)}{r_s^2}q^\mu-\frac{u_s^\mu}{r_s}
\right].
\end{equation}
We also collect all the total time derivatives involved in Harish-Chandra 
equation (\ref{HCme}). The term $m(\tau)u_\tau^\mu$ arises under the time 
derivative operator, where time-dependent function $m(\tau)$ is then nothing 
but the dynamical mass (\ref{mm}) of the particle. On rearrangement, the 
Harish-Chandra equation of motion (\ref{HCme}) coincides with the equation 
(\ref{me}) which is obtained via analysis of balance equations. It is in 
favour of the renormalization scheme for non-local theories developed in 
\cite{Yar3D,Y3D}.

To clear physical sense of the effective equation of motion (\ref{me}) we 
move the velocity term ${\dot m}u_\tau^\mu$ to the right-hand side of this 
equation:
\begin{equation}\label{meK}
m(\tau)a_\tau^\mu=\frac{g^2}{3}\left({\dot a}^\mu_\tau
-a^2_\tau u^\mu_\tau\right) + f_{\rm self}^\mu + f_{\rm ext}^\mu.
\end{equation}
According to \cite{Kos}, the scalar potential produces the Minkowski force
\begin{equation}
f_{\rm ext}^\mu = g\left(\eta^{\mu\alpha} + u_\tau^\mu 
u_\tau^\alpha\right)\frac{\partial\varphi_{\rm ext}}{\partial z^\alpha}
\end{equation}
which is orthogonal to the particle's 4-velocity. The self-force
\begin{equation}
 f_{\rm self}^\mu = g^2\int_{-\infty}^\tau {\rm d} sk_0^2
\frac{J_1(\xi)}{\xi}\left[
\frac{1+(q\cdot a_s)}{r_s^2}\left(q^\mu-r_\tau u_\tau^\mu\right) 
-\frac{u_s^\mu+(u_s\cdot u_\tau)u_\tau^\mu}{r_s}
\right]
\end{equation}
is constructed analogously from the tail part of gradient (\ref{grad}) of 
particle's own field (\ref{sg-sm}) supported on the world line $\zeta$. The 
own field contributes also to particle's inertial mass $m(\tau)$ defined by 
eq.(\ref{mm}).

\section{Conclusions}\label{Concl}
In the present paper, we find the radiative parts of energy-momentum and 
angular momentum carried by massive scalar field coupled to a point-like 
source. Scrupulous analysis of energy-momentum and angular momentum balance 
equations yields the  Harish-Chandra equation of motion of radiating scalar 
pole. This equation includes the effect of particle's own field as well as 
the influence of an external force.

To remove divergences stemming from the pointness of the particle we 
apply the regularization scheme originally developed for the case of  
electrodynamics in flat spacetime of three dimensions \cite{Yar3D,Y3D}. It 
summarizes a scrupulous analysis of energy-momentum and angular momentum 
carried by non-local electromagnetic field of a point electric charge. The 
simple rule allows us to identify that portion of the radiation which arises 
from source contributions interior to the light cone. 

Energy-momentum and angular momentum balance equations for radiating scalar 
pole constitute system of ten linear algebraic equations in variables 
$p^\mu_{\rm part}(\tau)$ and their first time derivatives ${\dot p}^\mu_{\rm 
part}(\tau)$ as the functions of particle's individual characteristics 
(velocity, acceleration, charge etc.). The system is degenerate, so that 
solution for particle's 4-momentum includes arbitrary scalar function, 
$m(\tau)$, which can be identified with the dynamical mass of the particle. 
Besides renormalization constant, the mass includes contributions from 
particle's own field as well as from an external field. 

This is a special feature of the self force problem for a scalar charge. 
Indeed, the time-varying mass arises also in the radiation reaction for a 
pointlike particle coupled to a massless scalar field on a curved background 
\cite{Q}. The phenomenon of mass loss by scalar charge is studied in 
\cite{BHPs,HPs}. Similar phenomenon occurs in the theory which describe a 
point-like charge coupled with massless scalar field in flat spacetime of 
three dimensions \cite{B}. The charge loses its mass through the emission of 
monopole radiation. 


\appendix

\section{Energy-momentum of the scalar massive field of uniformly moving 
particle}\label{unmvd}
\setcounter{equation}{0}
The simplest scalar field is generated by an unmoved source placed at the 
coordinate origin. Setting $z=(t,0,0,0)$ and $u=(1,0,0,0)$ in 
eq.(\ref{CMfi}), one can derive the static potential \cite{Bha,CM}:
\begin{equation}
\varphi (y)=g\frac{\exp(-k_0r)}{r}
\end{equation}
where $r=\sqrt{(y^1)^2+(y^2)^2+(y^3)^2}$ is the distance to the charge. 
It is the well-known Yukawa field.

In this Appendix we calculate the energy-momentum 
\begin{equation}\label{psc}
p_{\rm sc}^\nu(\tau)=\int_\Sigma {\rm d}\sigma_\mu T^{\mu\nu}
\end{equation}
carried by the scalar massive field due to a uniformly moving pointlike 
source $g$. The stress-energy tensor $\hat T$ is given by \cite{CM,Cw69,Cw70}
\begin{equation}
4\pi T_{\mu\nu}=\frac{\partial\varphi}{\partial y^\mu}
\frac{\partial\varphi}{\partial y^\nu}-\frac{\eta_{\mu\nu}}{2}
\left(
\eta^{\alpha\beta}\frac{\partial\varphi}{\partial y^\alpha}
\frac{\partial\varphi}{\partial y^\beta}+k_0^2\varphi^2
\right)
\end{equation}
and $\Sigma$ is an arbitrary space-like three-surface. 

It is convenient to choose the simplest plane $\Sigma_t=\{y\in{\mathbb 
M}_{\,4}:y^0=t\}$ associated with unmoving observer. We start with the 
spherical coordinates
\begin{equation}
y^0=s+r,\qquad y^i=rn^i
\end{equation}
where $n^i=(\cos\phi\sin\theta,\sin\phi\sin\theta,\cos\theta)$ and $s$ is 
the parameter of evolution. To adopt them to the integration surface 
$\Sigma_t$ we replace the radius $r$ by the expression $t-s$. On 
rearrangement, the final coordinate transformation $(y^0,y^1,y^2,y^3)\mapsto 
(t,s,\phi,\theta)$ looks as follows:
\begin{equation}
y^0=t,\qquad y^i=(t-s)n^i.
\end{equation}
The surface element is given by
\begin{equation}
{\rm d}\sigma_0=(t-s)^2{\rm d} s{\rm d}\Omega
\end{equation}
where ${\rm d}\Omega=\sin\theta{\rm d}\theta{\rm d}\phi$ is an element of 
solid angle.

After trivial calculation one can derive the only non-trivial component of 
energy-momentum (\ref{psc}) is 
\begin{eqnarray}
p^0_{\rm sc}&=&\frac{1}{4\pi}
\int_{-\infty}^t{\rm d} s(t-s)^2\int{\rm d}\Omega\frac12\left[
\sum_i\left(\frac{\partial\varphi}{\partial y^i}\right)^2+k_0^2\varphi^2
\right]
\\
&=&\frac{g^2}{2}\left[
k_0\exp[-2k_0(t-s)]+\frac{k_0\exp[-2k_0(t-s)]}{t-s}
\right]_{s\to-\infty}^{s\to t}
\nonumber\\
&=&\lim_{\varepsilon\to 0}\frac{g^2}{2\varepsilon} - \frac{g^2}{2}k_0
\nonumber
\end{eqnarray}
where $\varepsilon$ is positively valued small parameter. 

Having performed Poincar\'e transformation, the combination of 
translation and Lorentz transformation, we find the energy-momentum 
carried by massive scalar field of uniformly moving charge:
\begin{equation}\label{infin}
p^\mu_{\rm sc}=\lim_{\varepsilon\to 0}\frac{g^2}{2\varepsilon}u^\mu - 
\frac{g^2}{2}k_0u^\mu .
\end{equation}
The divergent Coulomb-like term is absorbed by the ``bare'' mass $m_0$ 
involved in action integral (\ref{Itot}) while the finite term contributes 
to the particle's individual 4-momentum (\ref{ppart}).

\section{Derivation of the dynamical mass}\label{mass}

The scalar product of particle 4-velocity on the first-order 
time-derivative of particle 4-momentum (\ref{pdot}) is as follows:
\begin{eqnarray}\label{udp}
({\dot p}_{\rm part}\cdot u_\tau)&=&\frac{g^2}{3}a_\tau^2+
\frac{g^2}{2}\int_{-\infty}^\tau {\rm d} s
k_0^2\frac{J_1(\xi)}{\xi}\left[
\frac{1+(q\cdot a_s)}{r_s^2}(q\cdot u_\tau)-\frac{(u_s\cdot u_\tau)}{r_s}+
\frac{(q\cdot a_\tau)}{r_\tau}
\right]\nonumber\\
&+&g\frac{{\rm d}\varphi_{\rm ext}}{{\rm d}\tau}.
\end{eqnarray}
Since $(u\cdot a)=0$, the scalar product of particle acceleration on the 
particle 4-momentum (\ref{ppart}) does not contain the scalar function 
$m$:
\begin{equation}\label{ap}
(p_{\rm part}\cdot a_\tau)=-\frac{g^2}{3}a_\tau^2-
\frac{g^2}{2}\int_{-\infty}^\tau {\rm d} s
k_0^2\frac{J_1(\xi)}{\xi}\frac{(q\cdot a_\tau)}{r_\tau}.
\end{equation}
Summing up (\ref{udp}) and (\ref{ap}) we obtain the non-local expression:
\begin{equation}\label{dpdp}
\frac{{\rm d}}{{\rm d}\tau}(p_{\rm part}\cdot u_\tau)=
\frac{g^2}{2}\int_{-\infty}^\tau {\rm d} s
k_0^2\frac{J_1(\xi)}{\xi}
\frac{\partial}{\partial s}\left[\frac{(q\cdot u_\tau)}{r_s}\right]
+g\frac{{\rm d}\varphi_{\rm ext}}{{\rm d}\tau}.
\end{equation}
We rewrite the expression under the integral sign as the following 
combination of partial derivatives in time variables:
\begin{equation}
k_0^2\frac{J_1(\xi)}{\xi}
\frac{\partial}{\partial s}\left(\frac{(q\cdot u_\tau)}{r_s}\right)=
\frac{\partial}{\partial s}\left(k_0^2\frac{J_1(\xi)}{\xi}
\frac{(q\cdot u_\tau)}{r_s}\right)-
\frac{\partial}{\partial \tau}\left(k_0^2\frac{J_1(\xi)}{\xi}\right).
\end{equation}
This circumstance allows us to integrate the expression (\ref{dpdp}) over
$\tau$:
\begin{eqnarray}
(p_{\rm part}\cdot 
u_\tau)&=&-m_0+\frac{g^2}{2}\int_{-\infty}^\tau{\rm d}\tau_1
\int_{-\infty}^{\tau_1}{\rm d}\tau_2\left[
\frac{\partial}{\partial \tau_2}\left(k_0^2\frac{J_1(\xi)}{\xi}
\frac{(q\cdot u_1)}{r_2}\right)-
\frac{\partial}{\partial \tau_1}\left(k_0^2\frac{J_1(\xi)}{\xi}\right)
\right]\nonumber\\
&+&g\varphi_{\rm ext}\nonumber\\
&=&-m_0-\frac{g^2}{2}\int_{-\infty}^\tau{\rm d} s
k_0^2\frac{J_1(\xi(\tau,s))}{\xi(\tau,s)}
+g\varphi_{\rm ext}.
\end{eqnarray}
To integrate the second term in between the square brackets we substitute 
$\int_{-\infty}^\tau{\rm d}\tau_2\int^{\tau}_{\tau_2}{\rm d}\tau_1$ for
$\int_{-\infty}^\tau{\rm d}\tau_1\int_{-\infty}^{\tau_1}{\rm d}\tau_2$.
The external potential is referred to the observation instant $\tau$.

Alternatively, the scalar product of 4-momentum (\ref{ppart}) and 
4-velocity is as follows:
\begin{equation}\label{pu}
(p_{\rm part}\cdot u_\tau)=-m+\frac{g^2}{2}\int_{-\infty}^\tau ds
k_0^2\frac{J_1[\xi(\tau,s)]}{\xi(\tau,s)}.
\end{equation} 
Having compared these expressions we obtain:
\begin{equation}\label{mas}
m=m_0+g^2\int_{-\infty}^\tau {\rm d} 
sk_0^2\frac{J_1[\xi(\tau,s)]}{\xi(\tau,s)}-
g\varphi_{\rm ext}.
\end{equation} 
We suppose that the renormalization constant $m_0$ already absorbs the 
Coulomb-like infinity which arises in eq.(\ref{infin}).

\end{document}